\documentstyle[12pt]{article}
\def\journal{\topmargin .3in	\oddsidemargin .5in
	\headheight 0pt	\headsep 0pt
	\textwidth 5.625in % 1.2 preprint size  %6.5in
	\textheight 8.25in % 1.2 preprint size 9in
	\marginparwidth 1.5in
	\parindent 2em
	\parskip .5ex plus .1ex		\jot = 1.5ex}
\journal

\catcode`\@=11
\def\marginnote#1{}
\newcount\hour
\newcount\minute
\newtoks\amorpm
\hour=\time\divide\hour by60
\minute=\time{\multiply\hour by60 \global\advance\minute by-\hour}
\edef\standardtime{{\ifnum\hour<12 \global\amorpm={am}%
	\else\global\amorpm={pm}\advance\hour by-12 \fi
	\ifnum\hour=0 \hour=12 \fi
	\number\hour:\ifnum\minute<10 0\fi\number\minute\the\amorpm}}
\edef\militarytime{\number\hour:\ifnum\minute<10 0\fi\number\minute}
\def\draftlabel#1{{\@bsphack\if@filesw {\let\thepage\relax
   \xdef\@gtempa{\write\@auxout{\string
      \newlabel{#1}{{\@currentlabel}{\thepage}}}}}\@gtempa
   \if@nobreak \ifvmode\nobreak\fi\fi\fi\@esphack}
	\gdef\@eqnlabel{#1}}
\def\@eqnlabel{}
\def\@vacuum{}
\def\draftmarginnote#1{\marginpar{\raggedright\scriptsize\tt#1}}
\def\draft{\oddsidemargin -.5truein
	\def\@oddfoot{\sl preliminary draft \hfil
	\rm\thepage\hfil\sl\today\quad\militarytime}
	\let\@evenfoot\@oddfoot	\overfullrule 3pt
	\let\label=\draftlabel
	\let\marginnote=\draftmarginnote
   \def\@eqnnum{(\theequation)\rlap{\kern\marginparsep\tt\@eqnlabel}%
\global\let\@eqnlabel\@vacuum}  }
\def\preprint{\twocolumn\sloppy\flushbottom\parindent 2em
	\leftmargini 2em\leftmarginv .5em\leftmarginvi .5em
	\oddsidemargin -.5in	\evensidemargin -.5in
	\columnsep .4in	\footheight 0pt
	\textwidth 10in	\topmargin  -.4in
	\headheight 12pt \topskip .4in
	\textheight 7.1in \footskip 0pt
	\def\@oddhead{\thepage\hfil\addtocounter{page}{1}\thepage}
	\let\@evenhead\@oddhead	\def\@oddfoot{}	\def\@evenfoot{} }
\def\numberbysection{\@addtoreset{equation}{section}
	\def\theequation{\thesection.\arabic{equation}}}
\def\underline#1{\relax\ifmmode\@@underline#1\else
	$\@@underline{\hbox{#1}}$\relax\fi}
\def\titlepage{\@restonecolfalse\if@twocolumn\@restonecoltrue\onecolumn
     \else \newpage \fi \thispagestyle{empty}\c@page\z@
	\def\thefootnote{\fnsymbol{footnote}} }
\def\endtitlepage{\if@restonecol\twocolumn \else \newpage \fi
	\def\thefootnote{\arabic{footnote}}
	\setcounter{footnote}{0}}  %\c@footnote\z@ }
\catcode`@=12
\relax
\def\figcap{\section*{Figure Captions\markboth
	{FIGURECAPTIONS}{FIGURECAPTIONS}}\list
	{Figure \arabic{enumi}:\hfill}{\settowidth\labelwidth{Figure 999:}
	\leftmargin\labelwidth
	\advance\leftmargin\labelsep\usecounter{enumi}}}
 \relax
\def\tablecap{\section*{Table Captions\markboth
	{TABLECAPTIONS}{TABLECAPTIONS}}\list
	{Table \arabic{enumi}:\hfill}{\settowidth\labelwidth{Table 999:}
	\leftmargin\labelwidth
	\advance\leftmargin\labelsep\usecounter{enumi}}}
 \relax
\def\reflist{\section*{References\markboth
	{REFLIST}{REFLIST}}\list
	{[\arabic{enumi}]\hfill}{\settowidth\labelwidth{[999]}
	\leftmargin\labelwidth
	\advance\leftmargin\labelsep\usecounter{enumi}}}
 \relax
\makeatletter
\newcounter{pubctr}
\def\publist{\@ifnextchar[{\@publist}{\@@publist}}
\def\@publist[#1]{\list
	{[\arabic{pubctr}]\hfill}{\settowidth\labelwidth{[999]}
	\leftmargin\labelwidth
	\advance\leftmargin\labelsep
	\@nmbrlisttrue\def\@listctr{pubctr}
	\setcounter{pubctr}{#1}\addtocounter{pubctr}{-1}}}
\def\@@publist{\list
	{[\arabic{pubctr}]\hfill}{\settowidth\labelwidth{[999]}
	\leftmargin\labelwidth
	\advance\leftmargin\labelsep
	\@nmbrlisttrue\def\@listctr{pubctr}}}
 \relax
\makeatother
\catcode`\@=11
\def\section{\@startsection {section}{1}{0pt}{-3.5ex plus -1ex minus
 -.2ex}{2.3ex plus .2ex}{\raggedright\large\bf}}
\catcode`\@=12
\newskip\humongous \humongous=0pt plus 1000pt minus 1000pt

\newif\ifdtup

\def\oldreffmt#1{\rlap{[#1]} \hbox to 2\parindent{}}

\def\figfmt#1{\rlap{Figure {#1}} \hbox to 1in{}}

\def\beq{\begin{equation}}
\def\eeq{\end{equation}}

\def\bea{\begin{eqnarray}}
\def\eea{\end{eqnarray}}
\catcode`\@=11
\def\eqnarray{\stepcounter{equation}\let\@currentlabel=\theequation
\global\@eqnswtrue
\global\@eqcnt\z@\tabskip\@centering\let\\=\@eqncr
\gdef\@@fix{}\def\eqno##1{\gdef\@@fix{##1}}%
$$\halign to \displaywidth\bgroup\@eqnsel\hskip\@centering
  $\displaystyle\tabskip\z@{##}$&\global\@eqcnt\@ne
  \hskip 2\arraycolsep \hfil${##}$\hfil
  &\global\@eqcnt\tw@ \hskip 2\arraycolsep $\displaystyle\tabskip\z@{##}$\hfil
   \tabskip\@centering&\llap{##}\tabskip\z@\cr}

\def\@@eqncr{\let\@tempa\relax
    \ifcase\@eqcnt \def\@tempa{& & &}\or \def\@tempa{& &}
      \else \def\@tempa{&}\fi
     \@tempa \if@eqnsw\@eqnnum\stepcounter{equation}\else\@@fix\gdef\@@fix{}\fi
     \global\@eqnswtrue\global\@eqcnt\z@\cr}

\catcode`\@=12
\font\tenbifull=cmmib10 % bold math italic
\font\tenbimed=cmmib10 scaled 800
\font\tenbismall=cmmib10 scaled 666
\relax

\def\thefootnote{\fnsymbol{footnote}}

\begin{document}
\begin{titlepage}
\begin{center}
April, 1992      \hfill     LBL-32107 \\
                    \hfill       UCB-PTH-92/08 \\
\vskip 0.2 in
{\large \bf A  Realistic Technicolor Model \\
 from 150 TeV down}\footnotetext{This work was supported in part by the
 Director, Office
of Energy Research, Office of High Energy and Nuclear Physics, Division of High
Energy Physics of the U.S. Department of Energy under contract
DE-AC03-76SF00098 and in part by the National Science Foundation under grant
PHY90-21139.}

\vskip .2 in
       {\bf  Raman Sundrum}
        \vskip 0.3 cm
       {\it Department of Physics\\
          University of California, Berkeley, CA 94720\\
          and\\
          Lawrence Berkeley Laboratory \\
          1 Cyclotron Road, Berkeley, CA 94720 \\   }
        \vskip 0.7 cm
%\vskip 0.3 in

\begin{abstract}
A realistic technicolor model is presented with the dynamics below $150$ TeV
treated explicitly. Electroweak symmetry is broken by the condensates of a
`minimal' doublet of technifermions. The new feature of the model is that the
the third generation quarks are unified with the technifermions into
multiplets of a walking gauge force down to a scale of $10$ TeV. The
remaining quarks and leptons are not involved in this unification however.
 The walking dynamics enhances the higher dimension
interactions which give the ordinary fermions their masses and mixing, while
leaving flavor-changing neutral currents suppressed. Because the third
generation quarks actually feel the walking force their masses can be much
larger than those of the other quarks and the leptons. The only non-standard
particles
with masses below several TeV are the single doublet of technifermions,
 so electroweak
radiative corrections are estimable and within experimental limits.

\end{abstract}
\end{center}

\end{titlepage}

\section{Introduction}

Models of dynamical electroweak (EW) symmetry breaking must satisfy several
stringent experimental and theoretical constraints in order to be realistic.
 First, they must include
interactions to communicate EW symmetry breaking to the ordinary fermions,
 strong enough to  give the observed masses
(in particular the top quark mass) and mixings, while
suppressing interactions which lead to unacceptably strong
flavor-changing neutral currents (FCNC's) \cite{thfcnc}. It is preferable
that this be done
without fine-tuning the couplings of the theory.  In this regard walking
technicolor (WTC) theories are a large step in the right direction \cite{walk}
since these theories contain  a dynamical enhancement of the interactions
responsible for fermion masses but not for those interactions leading to
FCNC's.
 Also, improving precision electroweak experiments have become sensitive to the
virtual effects of TeV-scale particles, thereby further restricting models of
the undiscovered Higgs sector \cite{prec}.

In this paper I will  outline a technicolor model,
 made explicit below a cutoff $\Lambda = 150$ TeV, which is consistent with the
present experimental  constraints. In this model
EW symmetry is broken by the condensates of a techni-doublet, $(T, B)$,
 just as in minimal technicolor (MTC) \cite{mtc}. Between $\Lambda$ and
and a scale $\mu_U = 9$ TeV the
 technicolor (TC) group is unified with the  color group of the
top and bottom quarks.  The resulting unified gauge force has
walking
gauge dynamics, due to the presence of several
heavy EW-singlet fermions  transforming under
the same gauge symmetry.  From the point
of view of the third generation quarks, this is a `walking extended
technicolor' theory as
contemplated by Georgi \cite{ggi}. The fact that the third generation quarks
feel the walking force leads to extra enhancement of their low-energy
interactions and masses beyond that which occurs in the standard WTC scenario,
where all the standard fermions
are singlets of the walking gauge group.

The extra walking enhancement is not required for the remaining quark and
lepton masses. In fact the first two generation quarks
 must not feel the walking  force
 because otherwise FCNC interactions would also be enhanced, as noted by
Georgi \cite{ggi}.  Therefore,  I take the  quarks of the first
two generations to
transform under
a separate color group from the top and bottom at high energies.
 Thus at high energies the third generation
quarks are more closely associated with the technifermions than with the
standard
fermions. At accessible energies this is not the case.
The transition is made at $\mu_U = 9$ TeV where a  new `ultra-color' (UC)
sector, becomes strongly coupled   and condensates of
UC-fermions form. The quantum number assignments of the UC-fermions are such
that the condensation breaks the various high-energy gauge groups down to  QCD
and TC. In order for this all to go through I must make the dynamical
assumption that
{\em gauge symmetries can be broken though their couplings
are not perturbatively weak} but are nevertheless below the critical coupling
necessary to trigger chiral symmetry breaking.

Due to the walking dynamics from $\Lambda$ down to $\mu_U$,
 large anomalous dimensions in the renormalization group (RG) flow
 enhance the necessary higher-dimension
operators. However, additional enhancement turns out to be
necessary in order to obtain  realistic  masses and mixings for
the quarks and leptons. This is provided by enhancing the low-energy effects of
a
four-technifermion operator beyond its natural strength.
 This requires tuning of the
corresponding
high-energy  coupling \cite{setc}. However, the tuning is only at
the ten percent level.

 The  walking effects are estimated using a  large-N
approximation, modified for a walking gauge theory.
In this approximation one can calculate the anomalous
dimensions of the important higher-dimension operators given the anomalous mass
dimension of fermions feeling the walking force.
(More details on
this type of RG analysis of walking gauge theories in the modified large-N
expansion will appear in Ref. \cite{me}.)

Most of the non-standard particles are very massive and decouple from  EW
physics at accessible energies. The exceptions are the technifermions which
therefore dominate the EW radiative
corrections from the non-standard physics. These effects are both
easily estimable and within experimental limits.

The cutoff in this effective theory, $\Lambda$, has been taken as large as
 the scale of the strongest four-fermion operator appearing in it.
 If the cutoff were  larger, the physics underlying the higher dimension
 operators would have to be described explicitly. However, the effective
 field theory language allows us to postpone the details of this
$\Lambda$-scale physics while focussing on the problem of dynamical EW
 symmetry breaking. The physics above $\Lambda$ may be supersymmetric, preonic,
 extended technicolor or something else.
     Recently, Einhorn and
Nash \cite{einh} have also
 discussed  technicolor theories compatible with the large top quark
mass from a similar viewpoint (though they stayed within  the purely
extended technicolor scenario).

The paper is organized as follows.
 In Section 2, the particle
content of the effective theory below $\Lambda$ is detailed.
In Section 3, I describe
the approximation in  which the dynamics between $\Lambda$
and $\mu_U$ is analysed.
  In Section 4,
the effective lagrangian is written down including the higher-dimension
operators which
communicate EW symmetry breakdown to the ordinary fermions, giving rise to
fermion masses and mixings.
 In Section 5, I define a quantitative
 measure of the tuning necessary to enhance the low-energy
 effects of the  four-technifermion coupling. Its value within the
 present model is given in Section 7.
 In Section 6, I describe the dynamics of the  UC sector which breaks the
 high energy gauge symmetries down
 to those of MTC.
In Section 7, the masses and mixings of the ordinary particles are estimated.
 In Section 8, the strengths of FCNC's due to  the non-standard physics
are estimated. In Section 9, EW radiative
corrections due to the technifermions are estimated. Section 10 provides the
conclusions of this paper.

\section{The model}

At the scale $\Lambda \sim 150$ TeV
the effective theory  has a gauge symmetry
\begin{equation}
SU(2)_{EW}  \times U(1)_{Y'} \times SU(3) \times SU(5)  \times SU(2)'
\times SU(M)_{UC} \times SU(M)'_{UC'}.
\end{equation}
$SU(2)_{EW}$ is the weak gauge group. $SU(3)$ is a weak color group carried by
the quarks of the first two generations. $SU(5)$ is the walking unified group
 for the three colors of the third generation quarks and two technicolors.
$SU(M)$ and
$SU(M)'$ are UC gauge groups. $SU(2)'$ is carried only by certain UC-fermions
in order to ensure the correct gauge symmetry breaking pattern at $\mu_U$. At
$\mu_U$  the weak $U(1)_{Y'}$ group combines with an $SU(5)$ generator to
give ordinary
hypercharge, $U(1)_Y$, QCD emerges, $SU(3)_{QCD} < SU(3) \times SU(5)$,
 and so does TC, $SU(2)_{TC} < SU(5) \times SU(2)'$.
 (Readers concerned about the vacuum alignment problem associated with TC
 theories with two technicolors \cite{vac} are assured that the matter will be
 satisfactorily resolved in Section 7.)

The model contains fermions belonging to four basic sectors. Each sector is
separately anomaly-free.

(i) Minimal Technicolor (MTC)

This sector includes all the ordinary particles of the standard model
 (minus the Higgs, of course)
and the technifermions which eventually break EW symmetry.  These particles
only  transform under the first four of the above
gauge groups, as
\begin{eqnarray}
(u, d)_L,~~ (c, s)_L & ~ &  (2, 3, 1)_{1/3} \nonumber \\
(u, d)_R,~~ (c, s)_R & ~ & (1, 3, 1)_{(4/3, -2/3)} \nonumber \\
L = (\nu_e, e)_L, ~~(\nu_{\mu}, \mu)_L, ~~(\nu_{\tau}, \tau)_L
& ~ & (2, 1, 1)_{-1} \nonumber \\
l_R = e_R,~~ \mu_R, ~~\tau_R & ~ & (1, 1, 1)_{-2} \nonumber \\
\psi_L = ({\cal T, B})_L = \pmatrix{\pmatrix{t \cr T},
\pmatrix{b \cr B}}_L & ~ &
(2, 1, 5)_{1/5} \nonumber \\
({\cal T, B})_R = \pmatrix{\pmatrix{t \cr T},  \pmatrix{b \cr B}}_R & ~ &
(1, 1, 5)_{(6/5, -4/5)}.
\end{eqnarray}
As can be seen, the technifermions $T, B$ are unified with the top and bottom
quarks at high energies, the three quark colors being the first three
components of each $5$ and the two technicolors being the last two components
of each $5$.
The first
two generations of quarks and all the leptons carry $Y'$ assignments given by
their standard model hypercharge assignments, while ${\cal T}$ and ${\cal B}$
have $Y'$
assignments designed to cancel anomalies and to make $t, T, b, B$ transform
in standard fashion under the hypercharge subgroup of the above gauge groups.
 The $Y'$ assignments have been written on the bottom-right as conventionally
done.   The quantum number assignments for the third generation quarks and the
technifermions are identical to those of the quarks and
 certain exotic fermions in the color-$SU(5)$ model of Foot and
Hernandez \cite{c5}. The assignment of
 different color groups to the third generation quarks and the quarks of the
 first two generations is similar to the construction of the
 `Topcolor' model of Chris Hill \cite{topc}.

(ii) Exotic fermions

These are heavy Dirac fermions which will  mix with $t$ and $b$ to a small
extent and
will allow interactions to emerge which will give rise to KM mixing between the
third generation quarks and the first two generation quarks.
 They transform under $SU(3) \times U(1)_{Y'}$ only, as

\begin{eqnarray}
(t', b')_L  & ~& 3_{(4/3, -2/3)} \nonumber \\
(t', b')_R  & ~& 3_{(4/3, -2/3)}.
\end{eqnarray}

(iii) Ultra-color (UC)

The purpose of this sector is to break $SU(3) \times SU(2') \times SU(5)
\times U(1)_{Y'}$ down to $SU(3)_{QCD} \times SU(2)_{TC} \times U(1)_{Y}$.
It is achieved by the QCD-like condensation of the UC fermions due to the
strong $SU(M)$ UC gauge force. These fermions transform under the full
gauge symmetry as
\begin{eqnarray}
{\cal Q}_L = \pmatrix{q \cr Q \cr}_L
&~~~~~~~~~~~~~~~~&(1, 1,  5, 1, M, 1)_{6/5} \nonumber \\
q_R &~~~~~~~~~~~~~~~~&(1, 3, 1,  1, M, 1)_{4/3} \nonumber \\
Q_R &~~~~~~~~~~~~~~~~&(1, 1, 1, 2', M, 1)_{1}, \nonumber \\
p_L &~~~~~~~~~~~~~~~~&(1, 3, 1, 1, 1, M')_{4/3} \nonumber \\
P_L &~~~~~~~~~~~~~~~~&(1, 1, 1, 2', 1, M')_{1},  \nonumber \\
{\cal P}_R = \pmatrix{p \cr P \cr}_R
&~~~~~~~~~~~~~~~~&(1, 1, 5, 1, 1, M')_{6/5}
\end{eqnarray}
 The $p$'s and $P$'s have been introduced
purely to cancel anomalies. They will be referred to as the UC$'$
sector, to distinguish them from the $q$'s and $Q$'s. We will take
 $M = M' = 3$ to
be specific but will continue to refer to $SU(M)$ and $SU(M)'$ in order to
distinguish them from  the  $SU(3)$ group already in the theory.
The basic structure of this sector of the theory is similar to that employed in
 Ref. \cite{luty}.

(iv) Extra $SU(5)$ fermions.

These I denote by $\chi$ and they are Dirac fermions, assumed to transform
 purely under $SU(5)$,
in such a way as to make it a walking gauge theory between $\Lambda$ and
$\mu_U$.  This is their sole
purpose in the model.

\section{The method of analysis}

I will now specify the approximation in which the RG flow between $\Lambda$ and
$\mu_U$ is treated, concentrating on providing a working set of rules to
capture
the effects of walking. Further details and a more careful discussion of the
assumptions made will appear in Ref. \cite{me}. Readers familiar with gap
equation analyses of WTC will find no surprises here.

The task at hand is to estimate the anomalous dimensions of the various
four-fermion operators which can appear in our effective lagrangian so that
their infrared enhancement can be calculated.
The walking $SU(5)$ dynamics will be analysed in the leading order of a  $1/N$
expansion capable of capturing the walking dynamics,
where $N = 5$ here. Such an expansion was suggested by Appelquist and
Wijewardhana \cite{walk} and first made use of in Ref. \cite{hsu}.
 The only modification of standard large-$N$ QCD
diagrammatics which
occurs in the walking theory is that $\chi$ loops are not suppressed because
there are many flavors of them or because they are in higher representations of
$SU(5)$. This is of course crucial in making the $SU(5)$ coupling walk and not
run.  I will assume that a suitable choice of $\chi$ representations exist so
that the $SU(5)$ coupling
walks over the hierarchy from $\Lambda$ down to $\mu_U$, and that the
anomalous mass dimension for fermions in the $5$ representation due to the
walking force is
\begin{equation}
\gamma \sim 0.8.
\end{equation}
The normalization of $\gamma$ is such that
$m(\mu)/m(\mu') = (\mu'/\mu)^{\gamma}$ for a running fermion mass
parameter $m$.
Gap equation analysis \cite{walk} shows that chiral symmetry breaking occurs
when the gauge coupling
is so strong that $\gamma > 1$, so we are assuming a gauge coupling which is
walking close to, but below  the critical value for chiral symmetry breaking.

Between $\Lambda$ and $\mu_U$  I will keep only  the effects of
 the walking force on
the RG flow of higher dimension operators, dropping the effects of the
(weaker) running gauge forces. The only exception is the UC force acting in the
 UC sector, which is presumed to be getting stronger in the infrared
 in QCD-like fashion so
 that it can break the high-energy gauge groups at $\mu_U$.  Also
 the $SU(5)$ coupling will be treated as approximately standing.
 It is clear that four-fermion operators which contain only fields which are
 $SU(5)$ singlets have zero anomalous dimension in this approximation.
 This leaves those operators
 with two or four fields transforming under $SU(5)$. It is convenient to think
 of these operators as being due to the exchanges of  heavy $SU(5)$-singlet
bosons. These auxiliary
fields have mass terms but no kinetic terms.
They act as a convenient book-keeping device. Depending on the
four-fermion operator the auxiliary boson is chosen to be a scalar or a
vector, whichever is the $SU(5)$-singlet.
 Thus the four-fermion operators can be rewritten as dimension-four
couplings of fermions to auxiliary scalar and vector fields.

The reason for doing this is that
the diagrams which dominate for large N are those in which the auxiliary fields
do not appear in loops. This implies that the anomalous dimensions of the
four-fermion operators are determined purely by the running of the fermion
couplings to the auxiliary fields, and the running of the auxiliary field
masses. Auxiliary scalars, $S$, have Yukawa
couplings, $\overline{f} S f$, while auxiliary vectors, $V_{\mu}$, couple to
fermionic currents as $V_{\mu} J_{\mu}$. The running due to the walking gauge
dynamics is now simple. If the $f$'s
are $SU(5)$-singlets neither type of coupling runs. If they are both $5$'s
then Yukawa couplings run with the mass anomalous dimension $\gamma$. However,
because $J_{\mu}$ is a conserved flavor current from the point of view of the
$SU(5)$ dynamics, the vector coupling  still does not run.

The masses for the auxiliary fields are naturally of order $\Lambda$ at that
scale, corresponding to four-fermion operators with strengths of order
$1/\Lambda^2$. At lower scales, interactions can renormalize these masses,
but these are $O(1)$ effects unless large cancellations occur.

The reader may wonder whether the auxiliary fields develop kinetic terms and
become truly propagating at lower energies. This actually does occur but the
particles are still very heavy and it is inessential to keep these kinetic
terms for the purposes of this paper. The properties of these kinetic terms are
 discussed in
Ref. \cite{tern} and will be further discussed in the present large-N RG
language in Ref. \cite{me}.

The rules outlined in this section will become clearer
  in the next section where they are
used to determine the effective lagrangian below the scale  $\Lambda$.

\section{The effective lagrangian}

The effective lagrangian at a scale $\mu$, $\mu_U < \mu < \Lambda$,  can be
written as
\begin{eqnarray}
{\cal L}(\mu)  =  {\cal L}_{MTC} + {\cal L}_{3-mix} +
{\cal L}_{P-mass}   + {\cal L}_{misc}.
\end{eqnarray}
The various parts have been named according to the role they play in our
story.
I will not write the kinetic and mass terms of the
fermions and the gauge fields explicitly because there is no mystery
concerning them. The only particles which
can have mass terms are the $\chi$'s and the $t'$ and $b'$.
 I will take the $\chi$ masses to be such
that they decouple from the theory once their job of maintaining
walking is done, namely at $\mu_U$. The
$t'$ and $b'$ are assigned masses of order $20$ TeV.
 Also I will omit all four-fermion operators
which remain completly unenhanced by the rules of the last section.  The
effects due to
these interactions  to which we are most sensitive are FCNC's, which are
examined separately in Section 8 and found to be acceptably weak. All other
four-fermion interactions which can occur will be explicitly shown.
I now explain the various parts of the lagrangian one at a time.

$\bullet$ ${\cal L}_{MTC}$:

Enhanced four-fermion interactions involving only the particles of the MTC
sector can be written in terms of the exchange of a single auxiliary scalar,
 $\phi$, which
transforms under the weak group, $SU(2)_{EW} \times U(1)_{Y'}$ as a $2_{1}$.
\begin{eqnarray}
{\cal L}_{MTC} &=& (\overline{U} ~\overline{D})_L~ y_D(\Lambda)~ \phi D_R
+ (\overline{U}~ \overline{D})_L~ y_U(\Lambda)~ \tilde{\phi} U_R
+ \overline{L} ~y_l(\Lambda)~ \phi l_R \nonumber \\
&+& (\Lambda/\mu)^{\gamma}
\overline{\psi}_L~ y_b(\Lambda) ~\phi {\cal B}_R
+ (\Lambda/\mu)^{\gamma}
\overline{\psi}_L ~y_t(\Lambda) ~\tilde{\phi} {\cal T}_R + h.c. \nonumber \\
 &-&  x(\mu) \Lambda^2
\phi^\dagger \phi,
\end{eqnarray}
 Here, $U$ and $D$ represent $(u, c)$ and $(d, s)$, so
that $y_D$ and $y_U$ are really $2 \times 2$ matrices in flavor space and $y_l$
is a three component vector (one for each charged lepton).
The enhancement factors follow the rules of the last section.

At $\mu = \Lambda$, integrating out $\phi$ yields a set of four-fermion
interactions with strengths of order $y^2/\Lambda^2$ if $x(\Lambda) = 1$.
 $x(\mu)\Lambda^2$ is the running mass-squared of $\phi$. As
mentioned in the last section, $x(\mu)$ is naturally of order one but can be
 {\em tuned} to be smaller. Its role is discussed in the next section. The
 $\phi$ kinetic terms induced below $\Lambda$ play no part in this paper and
 are therefore neglected.

  The interactions in ${\cal L}_{MTC}$ are responsible for
communicating the TC-induced EW breakdown to the quarks and leptons
in order to give
them their masses. In the language of the auxiliary scalar this proceeds
through $\phi$ first getting a non-zero vacuum expectation value.

$\bullet$ ${\cal L}_{3-mix}$:

Although $y_{D,U}$ can provide mixing between the quarks of the first two
generations through their off-diagonal elements, Yukawa couplings which can
produce mixing involving the third generation quarks are forbidden by gauge
invariance ($t, b$ are in $SU(5)$ multiplets while the other quarks are not).
This is where the exotic fermions, $t', b'$, come in. The important part of
 the effective lagrangian involving them is

\begin{eqnarray}
{\cal L}_{3-mix} &=& (\overline{U} ~\overline{D})_L~ y_{b'}(\Lambda)~ \phi b'_R
+ (\overline{U} ~ \overline{D})_L~ y_{t'}(\Lambda)~ \tilde{\phi} t'_R \nonumber
\\
&+& \frac{g_{b'}(\Lambda) (\Lambda/\mu)^{\gamma}}{\Lambda^2}
(\overline{b'}_L q_R) (\overline{\cal Q}_L {\cal B}_R) ~
+~ \frac{g_{t'}(\Lambda) (\Lambda/\mu)^{\gamma}}{\Lambda^2}
(\overline{t'}_L q_R) (\overline{\cal Q}_L {\cal T}_R)  \nonumber \\
&+& h.c.
\end{eqnarray}
I have written some of the above operators without auxiliary fields, but the
enhancement rules are obvious. $y_{b'}$ and $y_{t'}$ each have two components,
one for each of the first two generations.
Integrating out the heavy $t', b'$ fermions below their masses will lead to
operators which will permit mixing between the third and first two generation
quarks.

$\bullet$ ${\cal L}_{P-mass}$:

This part of the lagrangian is designed to give sizeable masses for
the lightest UC$'$ states so that they decouple from the low-energy theory.
I have not bothered to include the $O(1)$ $\Lambda$-scale coefficients of
 each operator. With this in mind,
\begin{equation}
{\cal L}_{P-mass} \sim
\frac{(\Lambda/\mu)^{\gamma}}{\Lambda^2} (\overline{{\cal P}}_R {\cal Q}_L)~
(\overline{q}_R p_L)
+ \frac{(\Lambda/\mu)^{\gamma}}{\Lambda^2} (\overline{{\cal P}}_R {\cal Q}_L)~
(\overline{Q}_R P_L),
\end{equation}
 The $SU(M)'$ UC$'$ force is taken to be considerably weaker than the UC force.
When the UC condensate forms the above operators induce TeV-scale `current
mass'
terms for the UC$'$ fermions.

$\bullet$ ${\cal L}_{misc}$:

Finally, the lagrangian contains various miscellaneous operators, most
compactly written in the form
\begin{eqnarray}
{\cal L}_{misc} &\sim& \frac{(\Lambda/\mu)^{2\gamma}}{\Lambda^2}
J_{\mu}^L J_{\mu}^R + \frac{(\Lambda/\mu)^{\gamma}}{\Lambda^2}
(\overline{t'}t')(\overline{\chi} \chi) \nonumber \\
&+& \frac{(\Lambda/\mu)^{\gamma}}{\Lambda^2}
(\overline{b'}b')(\overline{\chi} \chi),
\end{eqnarray}
where $J_{\mu}^L$ denotes any of the $SU(5)$-adjoint currents
 $\overline{\psi}_L
\gamma_{\mu} \psi_L, ~\overline{{\cal Q}}_L \gamma_{\mu} {\cal Q}_L,
  ~\overline{\chi}_L \gamma_{\mu} \chi_L$, and $J_{\mu}^R$ denotes any
of the $SU(5)$-adjoint currents  $\overline{{\cal T}}_R
\gamma_{\mu} {\cal T}_R,~ \overline{{\cal B}}_R
\gamma_{\mu} {\cal B}_R,~ \overline{{\cal P}}_R \gamma_{\mu} {\cal P}_R$,
$\overline{\chi}_R \gamma_{\mu} \chi_R$. ${\cal L}_{misc}$
 plays little part in our  model, although in Section 8 FCNC's induced by some
 of these operators will be estimated.

\section{Enhanced fermion masses by tuning four-fermion interactions}

 Four-fermion couplings at high energies can be tuned in
such a way as to enhance their effects at lower energies beyond what is
expected from
dimensional analysis. This phenomenon occurs even in the presence of walking
gauge dynamics and has been exploited in the past to obtain enhanced fermion
masses beyond the pure walking effect \cite{setc,chiv}.

In the present model, it will be necessary to tune the strength of the
operator,
$(\overline{\psi}_L {\cal T}_R)(\overline{{\cal T}_R}\psi_L)$,
 in order to give extra
 enhancement to fermion masses. It is convenient to continue using the
 auxiliary scalar language in discussing this issue.
Recall that $x(\mu)
\Lambda^2$ is the mass-squared of the auxiliary  scalar  at the scale
$\mu$. The dominant effect renormalizing this parameter is
 due to ${\cal T}$ loops in the auxiliary scalar's self-energy graphs.
 (${\cal T}$ loops dominate because they have the largest Yukawa couplings to
 $\phi$.)

The result is,
\begin{equation}
x(\mu) \Lambda^2 \sim \Lambda^2 -
\frac{5 (1 - (\mu/\Lambda)^{2(1-\gamma)})}{(1 - \gamma) 8 \pi^2}
y_t(\Lambda)^2  \Lambda^2.
\end{equation}
The $5$ is just the number of colors running around a loop. This
 estimate has been made in the RG approach \cite{me}. The basic result is
however simple to understand. Putting $\gamma = 0$ recovers the
one-loop result, which neglects the effects of  walking dynamics. The
first term is
just the `bare' scalar mass-squared ($x(\Lambda) = 1$).
 The dressing of the loop by $SU(5)$-gluons is an $O(1)$ effect.

Now, $y_t(\Lambda)$ can be tuned in order to make $x$ small and positive.
Because the four-fermion interactions in ${\cal L}_{MTC}$ are represented
by $\phi$ exchange this amounts to enhancing the strength of these interactions
at low energies, which in turn will help to enhance the masses of ordinary
fermions. Using the above equation as a rough guide, $x$ is tuned to be small
for $y_t^2 \sim 5$.

I will quantify the amount of tuning necessary by $x/x_{generic}$. The generic
choice would appear to be $x_{generic} \sim 1/2$, so $2 x$ is the measure of
tuning needed. In the end roughly a ten percent tuning will be required,
 $x\sim 0.05$.

\section{The effective theory below the UC scale}

  From $\Lambda$ down to $\mu_U$, walking dynamics  leads to
the effective lagrangian described in Section 4.
I assume that at this scale the running $SU(M)$ gauge coupling, taken
 to be much stronger
than the $SU(M)'$ coupling, rapidly gets strong and breaks the chiral
symmetries of the UC-fermions in QCD-like fashion. I will take the associated
 Goldstone boson decay constant, $f_U$, to be $\sim 7$ TeV.
 The resulting condensates are estimated from our experience with QCD as
\begin{equation}
\langle \overline{q} q \rangle \sim \langle \overline{Q} Q \rangle \sim
- 20 f_U^3.
\end{equation}
 I will suppose that the UC$'$ force is much weaker,
$f_U'\sim 1.5$ TeV.

The formation of the UC-condensate  breaks the full gauge symmetry of the
 model, Eq. (1), down to
\begin{equation}
SU(2)_{EW} \times U(1)_Y \times SU(3)_{QCD} \times SU(2)_{TC}.
\end{equation}
The MTC particles  now transform as
\begin{eqnarray}
(u, d)_L,~~ (c, s)_L, ~~ (t, b)_L & ~ &  (2, 3, 1)_{1/3} \nonumber \\
(u, d)_R,~~ (c, s)_R, ~~ (t, b)_R & ~ & (1, 3, 1)_{(4/3, -2/3)} \nonumber \\
L = (\nu_e, e)_L, ~~(\nu_{\mu}, \mu)_L, ~~(\nu_{\tau}, \tau)_L
& ~ & (2, 1, 1)_{-1} \nonumber \\
l_R = e_R,~~ \mu_R, ~~\tau_R & ~ & (1, 1, 1)_{-2} \nonumber \\
(T, B)_L  & ~ & (2, 1, 2)_{0} \nonumber \\
(T, B)_R  & ~ & (1, 1, 2)_{(1, -1)}.
\end{eqnarray}
 Hypercharge has emerged as a linear
 combination of the $Y'$ generator and a diagonal $SU(5)$ generator, so that
 for former members of $5$'s,
\begin{equation}
Y = Y' + 1/15~{\rm diagonal}(2, 2, 2,
- 3, - 3),
\end{equation}
and  for $SU(5)$-singlets, $Y = Y'$.
Thus the low-energy quantum numbers are indeed just those of  standard MTC.
The UC$'$ fermions now transform as
\begin{eqnarray}
p_{L,R} &~~~~~~~~~~~~~~~~&(1, 3, 1)_{4/3} \nonumber \\
P_{L,R} &~~~~~~~~~~~~~~~~&(1, 1, 2)_{1}.
\end{eqnarray}

The breaking of gauge symmetries makes $24$ gauge bosons massive, which means
that all $24$ UC Goldstone bosons  are eaten. One of these heavy
gauge bosons is a $Z'$ but it is so heavy (O(10 TeV) in mass) that
 its mixing with the ordinary $Z$ can be negelcted. The masses of the
gauge bosons are estimated by working to leading order in their gauge
couplings, which give  mass-squares $\sim \pi \alpha_5 f_U^2$.

 We will
therefore match effective lagrangians at $\mu_U^2 \sim \pi \alpha_5 f_U^2$.
Gap equation analyses suggest $\alpha_5 \sim 1/2$ \cite{walk}. Thus,
\begin{equation}
\mu_U = 9 \rm ~TeV.
\end{equation}
Below
this scale the effective lagrangian has the gauge symmetry Eq. (13), whereas
 above it the theory has the full gauge symmetry of Eq. (1). At this level of
approximation the matching of couplings reads
\begin{eqnarray}
\frac{1}{\alpha_{QCD}(\mu_U)} &\sim& \frac{1}{\alpha_{3}(\mu_U)} +
\frac{1}{\alpha_{5}(\mu_U)} \nonumber \\
\frac{1}{\alpha_{TC}(\mu_U)} &\sim& \frac{1}{\alpha_{2'}(\mu_U)} +
\frac{1}{\alpha_{5}(\mu_U)}.
\end{eqnarray}
At the matching scale QCD is weak. This is arranged by taking the $SU(3)$ force
to be weak.

Because the $P$'s feel TC, the masses of the light UC$'$ states must be known
before determining
the TC coupling at $\mu_U$. In the low-energy phase, the $p$'s and $P$'s are
permitted mass terms, and the interactions in ${\cal L}_{P-mass}$ provide them
upon UC condensation,
\begin{equation}
{\cal L}_{P-mass} \sim
\frac{(\Lambda/\mu)^{\gamma}}{\Lambda^2}~ 5 f_U^3~
\overline{p} p
+ \frac{(\Lambda/\mu)^{\gamma}}{\Lambda^2}~ 5 f_U^3~
 \overline{P}  P.
\end{equation}
One must decide what choice to make for $\mu$, the scale down to which the
walking enhancement occurs. This is not just $\mu_U$ now because the
${\cal Q}$'s have constituent masses $\sim 30$ TeV, scaling up from QCD. This
means that they effectively decouple at this scale, so $\mu \sim 30$ TeV is a
better estimate.
The induced ${\cal P}$  masses are then of order $1$ TeV.
 Because $f_U' = 1.5$ TeV, this means that even
the lightest UC$'$ states, the UC' Goldstone bosons, have masses $\sim \mu_U$.
Thus UC$'$ states do not affect the running of TC below $\mu_U$.
 Below this scale, the only fermions which feel TC
are $T$ and $B$.

I estimate that
\begin{equation}
\alpha_{TC}(\mu) \sim 1/3.
\end{equation}
 This follows by taking the TC Goldstone decay constant
 to be the weak scale, $f_T = 250$ GeV, and scaling up QCD data with the
appropriate modification due to the difference in the number of colors.
  From this estimate it follows that $\alpha_{2'}(\mu_U) \sim 1$.
This is a rather strong coupling but  below the critical coupling necessary
for the $SU(2')$ force to break the chiral symmetries of
the fermions which feel it, which is $\alpha_{2'}^{critical} \sim 3/2$ by gap
equation analysis \cite{walk}.

Since the $t'$ and $b'$ have masses $\sim 20$ TeV, these particles
should be integrated out at scales below their masses.
Integrating them out of the effective lagrangian and then proceeding down to
$\mu_U$ yields
\begin{eqnarray}
{\cal L}_{3-mix}(\mu_U) &=& \frac{g_{b'}(\Lambda) y_{b'}(\Lambda)
(\Lambda/\mu)^{\gamma}}{\Lambda^2 m_{b'}} (\overline{U}~\overline{D})_L
\phi~q_R~(\overline{\cal Q}_L {\cal B}_R) \nonumber \\
&+& \frac{g_{t'}(\Lambda) y_{t'}(\Lambda)
(\Lambda/\mu)^{\gamma}}{\Lambda^2 m_{t'}} (\overline{U}~\overline{D})_L
\phi~q_R~(\overline{\cal Q}_L {\cal T}_R) + ~h.c~+ ...
\end{eqnarray}
The ellipsis refers to other operators obtained upon integrating out the $t'$
and $b'$ but which are irrelevant at low energies. Once again the appropriate
choice for the scale below which walking enhancement is cut off is $\mu \sim
30$ TeV. Inserting the UC condensate,
for scales below $\mu_U$,
\begin{equation}
{\cal L}_{3-mix} \sim 0.02  g_{b'}(\Lambda) y_{b'}(\Lambda)
 (\overline{U}~\overline{D})_L
\phi~b_R + 0.02 g_{t'}(\Lambda) y_{t'}(\Lambda)
 (\overline{U}~\overline{D})_L
\phi~t_R.
\end{equation}
These Yukawa couplings mix the first two generation quarks with the third as
required.

  Only the MTC particles remain in the theory below $\mu_U$ with a set of
four-fermion interactions mediated by $\phi$ exchange
which I will now show lead to  masses and mixings for the quarks and
leptons.

\section{Masses and couplings of the MTC particles}

The technifermions condense, break EW symmetry and are
confined in standard fashion. With more than two technicolors this statement
would require no explanation, but there is a subtlety when there are only two
technicolors since the doublet representation of $SU(2)_{TC}$ is
pseudo-real. This implies that the $T$ and $B$ give rise to an $SU(4)$ chiral
flavor symmetry for the pure TC theory instead of the familiar
 $SU(2) \times SU(2)$.
TC breaks the $SU(4)$ symmetry down to $Sp(4)$ instead of $SU(2)_V$. Five
 Goldstone bosons result instead of three. Suppose that the only forces present
 in the theory were EW and TC. Then one can show that in the standard model
 vacuum,
 where $U(1)_{EM}$ is the preserved symmetry, three of the Goldstone bosons
 would be eaten but the remaining two Goldstone bosons would have {\em
 negative}
mass-squares
 due to EW forces $\sim - (100~ {\rm GeV})^2$ \cite{vac}. This shows that the
standard model vacuum is not the
 right one. If this were the case it would be a phenomenological disaster.
 Fortunately there are relatively strong four-technifermion  forces in the
present model which stabilize the
 standard model vacuum. For the moment the reader should accept that the
 correct EW breaking pattern arises and I will show that this choice of vacuum
 is stable later.

The dominant mass contributions to MTC particles come from their four-fermion
couplings to the technifermions.
In the language where the auxiliary scalar field $\phi$ is present, these
contributions are
written in terms of
$\langle \phi \rangle$, which develops
as the result of TC condensation. This quantity is found by
 extremizing the
dominant terms in the effective potential,
\begin{equation}
V_{eff}(\phi) \sim  -
(\Lambda/\mu_U)^{\gamma} y_b(\Lambda)
\langle \overline{\psi}_L   {\cal B}_R \rangle \phi
-  (\Lambda/\mu_U)^{\gamma} y_t(\Lambda)  \langle
\overline{\psi}_L   {\cal T}_R \rangle \tilde{\phi} + \rm h.c.
  ~+ x \Lambda^2 \phi^{\dagger} \phi
\end{equation}
Note that the walking enhancement ceases when $SU(5)$ is broken at $\mu_U$.
Scaling up from QCD, and accounting for the difference in the number of colors,
\begin{equation}
\langle \overline{T} T \rangle \sim \langle \overline{B} B \rangle \sim
- 25 f_T^3.
\end{equation}
Using the fact that $y_b \ll y_t$,
\begin{equation}
\langle \phi \rangle \sim \frac{y_t(\Lambda)}{x} 100 ~\rm MeV.
\end{equation}
Substituting $\langle \phi \rangle$ into ${\cal L}_{MTC}$ gives a set of mass
terms for the fermions. Notice that fermion masses will be enhanced because of
the enhancement of four-fermion operators as measured by $1/x$.

 There is an extra  contribution to the masses of the
third generation quarks due to the  operators arising from the exchange
of massive  $SU(5)$ gauge bosons which link $t$ with $T$ and $b$ with $B$.
Such operators are not described by $\phi$ exchange. Estimating the operators
by one-SU(5)-gluon exchange,
this `extended technicolor' contribution follows in standard fashion. It
gives a common mass  to the $t$ and $b$ quarks,
\begin{eqnarray}
m_{ETC} &\sim& 2 \langle \overline{T} T \rangle /f_U^2 \nonumber \\
&\sim& 15~ \rm GeV.
\end{eqnarray}

Before attempting to pick couplings in order to obtain a fully realistic
spectrum, let us see what the generic prediction is. That is,  take
$y_{f}(\Lambda) \sim 1$ for all fermions, $f$, so that  $x \sim 1/2$.
 Then $\langle \phi \rangle \sim 400$ MeV.
Referring to the expression for ${\cal L}_{MTC}$ in Section 4, and taking into
account the extra ETC mass contribution for the third generation quarks, one
obtains
\begin{eqnarray}
m_{t,b} &\sim& 20 \rm~GeV \nonumber \\
m_{other} &\sim& 200 \rm ~MeV.
\end{eqnarray}
Using Eq. (21), the mixing between the third generation quarks and the other
quarks is given by a mixing mass
\begin{equation}
m_{3-mix} \sim 10 ~\rm MeV.
\end{equation}
This is in fact a decent caricature of the observed particle spectrum. There is
a large hierarchy between the masses of
the third generation quarks and the other quarks and leptons, along with small
mixing angles involving the third generation.

Now I will choose couplings in order to get a realistic spectrum.
  Third generation quark masses are
given by
\begin{equation}
m_{t,b} = (\Lambda/\mu_U)^{\gamma} y_{t,b}(\Lambda) \langle \phi \rangle
+ m_{ETC}.
\end{equation}
A large top mass requires that $x$ be tuned to be small in order to enhance
$\langle \phi \rangle$ sufficiently. In order to get a mass of $100$ GeV, I
estimate that, $x \sim 0.05$, $y_t(\Lambda)^2 \sim 5$.
This is the necessary ten percent tuning discussed  in Section 5.
  With this choice,
\begin{equation}
\langle \phi \rangle \sim 4.5 \rm~GeV.
\end{equation}
Thus $y_b(\Lambda) \sim - 1/5$, in order to get the observed bottom quark mass.
The remaining quark and lepton masses and mixing masses are of the form
\begin{equation}
m_f = y_f(\Lambda) \langle \phi \rangle,
\end{equation}
so they can all be realistically obtained with  $y_f \leq 1/2$.

We can also associate  `current' masses for the technifermions $T, B$
due to their Yukawa couplings to $\phi $, in ${\cal L}_{MTC}$. These couplings
are  identical to those for $t$
and $b$, by  $SU(5)$-invariance. Therefore we find that
\begin{eqnarray}
m_T^{current} \sim 100~ {\rm GeV} \nonumber \\
m_B^{current} \sim 10~  {\rm GeV}.
\end{eqnarray}
This will provide the major effect in estimating the non-standard
contribution to
the $T$ parameter of Peskin and Takeuchi \cite{prec}. These current masses
also represent the crucial vacuum stabilizing effect of the four-technifermion
interactions. In
particular one can use chiral perturbation theory to estimate the contribution
they make to the mass-squares of  the two pseudo-Goldstone bosons (PGB's)
 mentioned
at the beginning of this section. The result is $\sim +(500~ {\rm GeV})^2$.
This completly overwhelms the EW forces which are trying to destabilize the
standard model vacuum as represented by their negative mass-square
contributions
to the PGB's, so disaster has been averted.

  From Eq.(21), the typical size of the mixing masses involving the third
generation  is
\begin{eqnarray}
m_{3-mix} &\sim& 0.02 g_{b', t'}(\Lambda) y_{b', t'}(\Lambda)
\langle \phi \rangle
\nonumber \\
&\sim& g_{b', t'}(\Lambda) y_{b', t'}(\Lambda) 100~\rm MeV.
\end{eqnarray}
For $g$'s of order one this gives  mixing of the
correct size.

\section{Flavor-changing neutral currents}

To begin, consider FCNC's due to four-fermion operators that
can occur in the effective lagrangian at $\Lambda$. Because of the different
quantum numbers carried by the third generation quarks, these $\Lambda$-scale
 flavor-changing operators can only involve the first two generations.
At this scale, quarks from the first two generations, $f$, must
have four-fermion couplings at most as strong as $y_f/\Lambda^2$, as
can be seen by integrating $\phi$ out. Following the analysis of Dimopoulos
and Ellis \cite{thfcnc} this suggests that flavor-changing four-fermion
 operators, such as
$\overline{s}d~\overline{s}d$ will
naturally be present with  (Cabibbo suppressed) strengths of order
$y_f/(20 \Lambda^2)$.  These operators are not
enhanced by walking dynamics. Recall that $\langle \phi \rangle \sim 4.5$ GeV,
 so
$y_s \sim 0.05$. Thus $\overline{s}d~\overline{s}d$ is suppressed in the
effective lagrangian by $1/(3000 ~{\rm TeV})^2$. This is phenomenologically
acceptable. Other flavor-changing operators are  less constrained
by experiment.

If we  turn off the interactions which give mixing involving the third
generation (${\cal L}_{3-mix} \rightarrow 0$)
the non-standard physics below $\Lambda$ does not give rise
to new FCNC's. FCNC's develop because the third generation quarks couple
differently to the non-standard sectors than the first two generations
 (${\cal L}_{misc}$ and $SU(5)$-gauge-boson exchange) and when
we rotate from the gauge basis to the mass basis for the quarks these couplings
which were flavor diagonal in the gauge basis become  slightly off-diagonal, so
heavy non-standard states can mediate FCNC's. Therefore the new FCNC's involve
the third generation quarks, and are suppressed by the small third generation
mixing angles and the masses of the non-standard states, which are several TeV.
 This suggests for example that
$\overline{b}s~\overline{b}s$ has a strength {\em at most}
$\sim V_{bc}^2/( {\rm 10 TeV})^2$,
$\overline{b}d~\overline{b}d$ has a strength at most
$\sim V_{bu}^2/(10 {\rm TeV})^2$ and
$\overline{b}s~\overline{\mu}\mu$ has a strength at most
$\sim V_{bc}/(10 {\rm TeV})^2$. These rough upper bounds
  are below
the experimental limits on $\Delta B \neq 0$ FCNC's \cite{fcnc}.

\section{Electroweak radiative corrections}

The dominant non-standard contributions to the $S$ and $T$ parameters of Peskin
and Takeuchi \cite{prec} come from the weak scale TC sector, which we have
 seen is just
MTC. Peskin and Takeuchi \cite{pt} have given  the estimates for MTC
with two technicolors
(where the $T - B$ mass splitting
is given by $\sim m_t$, as in the present model),
\begin{eqnarray}
S \sim 0.22 \nonumber \\
T \sim 0.3.
\end{eqnarray}
 The values above lie  within the $90$ percent confidence level ellipse in
the experimental $S-T$ plane \cite{pt}.

One may wonder about the effects of the two PGB's on $S$, not taken into
account in the formula above. The answer is that virtual pairs of these
 PGB's {\em do not} contribute to the EW gauge boson mixing parameter $S$
because
 PGB-pair couplings to EW currents are proportional to their custodial
$SU(2)_{cust}$ quantum numbers, but the PGB's in this model are singlets under
 $SU(2)_{cust}$.

\section{Conclusion}

I have constructed a realistic technicolor model of EW symmetry breaking,
effective below $150$ TeV. It uses a combination of walking dynamics and strong
four-fermion interactions to  enhance
operators responsible for giving the ordinary fermions their masses, while
leaving operators responsible for FCNC's unenhanced. For the leptons and
quarks of the first
two generations this is very similar to the schemes already in the literature.
 However the third
generation quarks are treated differently in this model because they feel the
walking force. This gives extra enhancement to their interactions and to the
final top quark mass. At an ultra-color scale of several TeV, the theory
becomes
identical to minimal technicolor, with  the desired set of four-fermion
interactions emerging from the high energy walking dynamics. Mixing with the
third generation quarks is naturally small because they feel a different high
energy color force from the first two generations. Mixing then neccessarily
proceeds through very high dimension operators, produced in the present
model  by integrating out some heavy fermions.
 The minimal low-energy content of the theory results in electroweak
radiative corrections which are estimable,  moderate and phenomenologically
 acceptable.

\section*{Acknowledgements}

I am grateful to Markus Luty for reading the manuscript of the paper several
times and for many useful criticisms and also to Riccardo Rattazzi for
 discussions related to
various aspects of this paper. Uri Sarid introduced me to the Color-$SU(5)$
model. This work was supported in part by the Director, Office
of Energy Research, Office of High Energy and Nuclear Physics, Division of High
Energy Physics of the U.S. Department of Energy under contract
DE-AC03-76SF00098 and in part by the National Science Foundation under grant
PHY90-21139.

%\section*{Appendix}


\newpage

\begin{thebibliography}{99}

\bibitem{thfcnc} E. Eichten and K. Lane, Phys. Lett. 90B (1980) 125; S.
 Dimopoulos
 and J. Ellis, Nucl. Phys. B182 (1981) 505.

\bibitem{walk} B. Holdom, Phys. Lett. 150B (1985) 301;
V.A. Miransky, Nuovo Cimento 90A
(1985) 149; T. Appelquist, D. Karabali
and
L.C.R. Wijewardhana, Phys. Rev. Lett. 57 (1986) 957;
 M. Bando, T. Morozumi, H. So
and
K. Yamawaki, Phys. Rev. Lett. 59 (1987) 389;
T. Appelquist and L.C.R. Wijewardhana, Phys. Rev. D36 (1987) 568.

\bibitem{prec} R. Renken and M. Peskin, Nucl. Phys. B211  93 (1983);
B.W. Lynn, M.
Peskin and R.G. Stuart in `Physics at LEP', eds. J. Ellis and R. Peccei (1986);
 D. Kennedy and B.W. Lynn, Nucl. Phys. B322 (1989) 1;  M. Golden and L.
Randall,
Nucl. Phys. B361, 3 (1991); B. Holdom and J. Terning, Phys.
Lett. 247B (1990) 88; M. Peskin and T. Takeuchi, Phys. Rev. Lett. 65 (1990)
964;
 A. Dobado, D. Espriu and M. Herrero, Phys. Lett. 255B
(1990) 405; W.J. Marciano and J.L. Rosner, Phys. Rev. Lett. 65 (1990) 2963;
 R. Johnson, B.L. Young and D.W. McKay, Phys. Rev. D43 (1991) R17.

\bibitem{mtc}  S. Weinberg,
Phys. Rev. D19 1277 (1979); L. Susskind, Phys. Rev. D20 2619 (1979).

\bibitem{ggi} H. Georgi, Talk presented at SCGT90, Nagoya, HUTP-90-A048 (1990).

\bibitem{setc} T. Appelquist, M. Einhorn, T. Takeuchi and L.C.R. Wijewardhana,
Phys. Lett. B220 223 (1989); V.A. Miransky and K. Yamawaki, Mod. Phys. Lett. A4
129 (1989); V.A. Miransky, M. Tanabashi and K. Yamawaki, Phys. Lett. B221 177
(1989), Mod. Phys. Lett. A4 1043 (1989); T. Takeuchi, Phys. Rev. D40 2697
(1989); T. Appelquist and O. Shapira, Phys.
Lett. B249 83 (1990).

\bibitem{me} R. Sundrum, in preparation.

\bibitem{einh} M.B. Einhorn and D. Nash, NSF-ITP-91-91 (1991).

\bibitem{vac} M.E. Peskin, Nucl. Phys. B175 197 (1980); J.P. Preskill, Nucl.
Phys. B177 21 (1981).

\bibitem{c5} R. Foot and O.F. Hernandez, Phys. Rev. D41 2283 (1990).

\bibitem{topc} C.T. Hill, Phys. Lett. B266 419 (1991).

\bibitem{luty} M.A. Luty, LBL-32089 (1992).

\bibitem{hsu} R. Sundrum and S.D.H. Hsu, LBL-31066, UCB-PTH-91/34.

\bibitem{tern} T. Appelquist, J. Terning and L.C.R. Wijewardhana, Phys. Rev.
D44 (1991).

%\bibitem{tcon} Y. Nambu, EFI-88-39 (1988), EFI-88-62 (1988), EFI-89-08 (1989);
%V.A. Miransky, M. Tanabashi and K. Yamawaki, Mod. Phys. Lett. A4 1043 (1989),
%Phys. Lett. B221 177 (1989); W. Marciano, Phys. Rev. Lett. 62 2793 (1989);
% W.A.
%Bardeen, C.T. Hill and M. Lindner, Phys. Rev. D41 1647 (1990).

%\bibitem{nsh} T. Appelquist, U. Mahanta, D. Nash and L.C.R. Wijewardhana,
%Phys. Rev. D43 R646 (1991).

\bibitem{chiv} R.S. Chivukula, A. Cohen, K. Lane, Nucl. Phys. B343 554 (1990).

\bibitem{fcnc}  D. Wyler, in Proceedings of the XXIII International Conference
on High Energy Physics, Berkeley (1986).

\bibitem{pt} M.E. Peskin and T. Takeuchi, SLAC-PUB-5618 (1991).



\end{thebibliography}
\end{document}